\documentclass[aps,prc,nofootinbib,twocolumn,superscriptaddress,showpacs,floatfix]{revtex4}
\usepackage{mathrsfs}
\usepackage{latexsym}
\usepackage{amsmath}
\usepackage{amssymb}
\usepackage{graphicx}
\usepackage{longtable}

\usepackage{bbm}
\usepackage{epsfig}
\usepackage[usenames]{color}
\usepackage[colorlinks=true,citecolor=blue,linkcolor=blue]{hyperref}

\begin{document}

%\begin{CJK*}{GBK}{song}

%\begin{CJK*}{GB}{}

\title{Influence of vector interactions on the hadron-quark/gluon phase transition}
\author{G.Y. Shao}
%\email[Corresponding author: ]{shaogy@pku.edu.cn}
\affiliation{INFN-Laboratori Nazionali del Sud, Via S. Sofia 62, I-95123
Catania, Italy}
%\affiliation{Department of Physics, Henan Normal university, Xinxiang 453007, China}

\author{M. Colonna}
\email[Corresponding author. \\  Email: shaogy@pku.edu.cn, colonna@lns.infn.it, ditoro@lns.infn.it, 
liub@mail.ihep.ac.cn, matera@fi.infn.it]{}
\affiliation{INFN-Laboratori Nazionali del Sud, Via S. Sofia 62, I-95123
Catania, Italy}

\author{M. Di Toro}
\affiliation{INFN-Laboratori Nazionali del Sud, Via S. Sofia 62, I-95123
Catania, Italy}
\affiliation{Physics and Astronomy Dept., University of Catania, Via S. Sofia 64, I-95125 Italy}

\author{B. Liu}
\affiliation{IHEP, Chinese Academy of Sciences, Beijing, 100049
China} \affiliation{Theoretical Physics Center for Scientific
Facilities, Chinese Academy of Sciences, Beijing, 100049 China}

\author{F. Matera}
\affiliation{Physics and Astronomy Dept., University of Florence }
\affiliation{INFN-Sezione di Firenze, 
Via G. Sansone 1, I 50019 Sesto F.no (Firenze), Italy}

%\author{Y.X. Liu}
%\affiliation{Department of Physics and State Key Laboratory of \\
%Nuclear Physics and Technology,
%Peking University, Beijing 100871, China}
%\affiliation{Center of Theoretical Nuclear Physics,\\ National Laboratory of
%Heavy Ion Accelerator, Lanzhou 730000, China}

%\date{\today}

\begin{abstract}
The hadron-quark/gluon phase transition is studied in the two-phase
model. As a further study of our previous work, both the isoscalar
and isovector vector interactions are included in the Polyakov loop
modified Nambu--Jona-Lasinio model~(PNJL) for the quark phase. The relevance of 
the exchange ( Fock ) terms is stressed and suitably accounted for. The
calculation shows that the isovector vector interaction delays the phase
transition to higher densities and the range of the mixed phase
correspondingly shrinks. Meanwhile the asymmetry parameter of quark matter 
in the mixed phase decreases with the strengthening of this interaction 
channel. This 
%, which 
leads to some possible observation signals being
weakened, although still present. We show that these can be 
%are 
rather general effects of a repulsion in the quark phase due to the 
symmetry energy. This is also   
%of a symmetry repulsion in the quark sector,
confirmed by a simpler calculation with the MIT--Bag model.
%in simpler Hadron-MIT-Bag calculations. 
However, the asymmetry parameter of quark matter is slightly enhanced with the
inclusion of the isoscalar vector interaction, but the phase
transition will be moved to higher densities. The largest
uncertainty on the phase transition lies in the undetermined
coupling constants of the vector interactions. 
In this respect new data on the mixed phase obtained from Heavy Ion 
Collisions at Intermediate Energies appear very important.

\end{abstract}


\maketitle

\section{Introduction}
The phase transition from nuclear matter to quark-gluon matter
is one of the most concerned topics in modern physics related
to heavy-ion collision experiments and compact stars.
As a principle tool, lattice QCD provides us a framework to
investigate non-perturbative phenomena, such as confinement
and quark-gluon plasma formation at finite temperature and
vanishing~(small) chemical potential
~\cite{Karsch01, Karsch02, Allton02, Kaczmarek05, Cheng06, YAoki99, Borsanyi10}.
However, lattice calculations suffer the sign problem at large
chemical potential. To evade this problem several approximation
methods have been proposed~\cite{Fodor02,Fodor03,Elia09,Ejiri08,Clark07},
but the validity  of the result at $\mu_q/T>1$ still should be taken
with care~\cite{Fukushima11}. On the other hand, to give a complete
description of  QCD phase diagram some phenomenological effective models~\cite{Nambu61,Toublan03,Werth05,Abuki06,Volkov84,Hatsuda84,Klevansky92,Hatsuda94,
Alkofer96,Buballa05, Rehberg95} have been also developed. Among
these models, the Nambu--Jona-Lasinio~(NJL) type models~\cite{Hatsuda84,Klevansky92,Hatsuda94,
Alkofer96,Buballa05, Rehberg95}, especially those coupled with Polyakov loop~(PNJL)~\cite{Fukushima04,Ratti06,Costa10,Schaefer10,Herbst11,Kashiwa08,Abuki08,Fu08}
are predominant, offering a simple illustration of chiral symmetry
breaking and restoration, as well as (de)confinement effect.

The lattice QCD and (P)NJL type models are based on the degrees of
freedom of quarks and gluons. Recently, the two-phase model  with
both hadron and quark degrees of freedom, widely used in the
description of the phase transition in neutron star matter under the
weak equilibrium~(e.g.,\,\cite{Glendenning92,Glendenning98,
Burgio02,Maruyama07,Yang08,Shao10,Shao110,Xu10,Dexheimer101,Dexheimer102}
), has also been taken to study the phase transition related to
heavy-ion collisions~\cite{Muller97, Toro06,Toro09, Torohq11, Liu11,
Cavagnoli10,Pagliara10,Shao111,Shao112}, particularly the phase
transition in asymmetric matter.
The latter is
possible to be probed in the planned facilities, such as FAIR at
GSI-Darmstadt and NICA at JINR-Dubna~\cite{Toro06, Torohq11,
Pagliara10, Cavagnoli10,Shao111,Shao112,Liu11}.
In these studies, only the scalar
interacting channel was considered for quark matter, and this
channel interaction is responsible for the dynamical masses of
quarks. The isoscalar-vector channel interaction was also included
to study the properties of quark matter~\cite{Costa10,Carignano10}
or the hadron-quark phase transition in neutron star
matter~\cite{Blaschke10,Ohnishi11}.

However, up to date, the vector interacting channels (including both
the isoscalar vector and isovector vector interaction) have not been
considered in describing quark matter, in the context of heavy-ion collisions,
in the two-phase model. The inclusion of these vector interactions
will modify the quark pressure and chemical potentials. In particular,
the isovector-vector interacting channel contributes to $u,\,d$
quark flavors differently. Correspondingly, the onset densities of
the quark phase are possibly modified and some observation signals of
hadron-quark phase transition in asymmetric matter may be
influenced. Therefore, it is important to study the effect of the vector
channel interactions  on the hadron-quark phase transition, and
related information deduced from new data in heavy ion collisions
at intermediate energies.

The paper is organized as follows. In Section II, we describe
briefly the modified effective quark model and give the relevant
formulas with the newly included isovector and isoscalar vector
interactions. In Section III, we present the numerical results and
discuss the influence of vector channel interactions on the
hadron-quark phase transition of dense asymmetric matter. For the
isovector part we include also some results with an Isospin-MIT-Bag
model, just to stress the physics behind the symmetry terms in the
quark sector, not dependent on the models. Finally, a summary is
given in Section IV.

\section{ The model}\allowdisplaybreaks
In the two-phase model hadron matter  is
described by the nonlinear Walecka model, and quark matter
is described by the PNJL model with the newly added vector interactions.
%The mixed phase between 
In the mixed phase, 
the pure hadronic phase and quark phase
are connected to each other through the Gibbs conditions with the
thermal, chemical and mechanical equilibria,
based on baryon number and isospin conservation in the strong
interacting process.

The Relativistic Mean Field (RMF) approach will be taken to describe
the properties of hadronic matter. This model can provide an
excellent description of  nuclear matter and finite nuclei. The
exchanged mesons in this model include the isoscalar-scalar
meson~($\sigma$), isoscalar-vector meson~($\omega$),
isovector-vector meson~($\rho$) and  isovector-scalar
meson~($\delta$). This is called the Non Linear-$\rho,\delta$
($NL\rho\delta$) effective interaction. For details, see
Refs.~\cite{Shao10,Toro06,Toro09,Torohq11,Liu11} and references
therein.

For the quark phase, we adopt an extension of the two-flavor NJL model
to include the Polyakov loop contribution\cite{Fukushima04,Ratti06}. The Lagrangian
is given by
\begin{eqnarray}
\mathcal{L}_{q}&=&\bar{q}(i\gamma^{\mu}D_{\mu}-\hat{m}_{0})q+
G_\sigma\bigg[(\bar{q}q)^{2}+
(\bar{q}i\gamma_{5}\vec{\tau} q)^{2}\bigg]\nonumber \\
&&+G_\delta\bigg[(\bar{q}\vec{\tau} q)^{2}+
(\bar{q}i\gamma_{5}q)^{2}\bigg]\nonumber\\
&&-G_\rho\bigg[(\bar{q}\gamma^\mu \vec{\tau} q)^{2}+
(\bar{q}\gamma_{5}\gamma^\mu \vec{\tau} q)^{2}\bigg]\nonumber\\
&&-G_\omega\bigg[(\bar{q}\gamma^\mu q)^{2}+
(\bar{q}\gamma_{5}\gamma^\mu q)^{2}\bigg]\nonumber \\
&&-\mathcal{U}(\Phi[A],\bar{\Phi}[A],T)
\end{eqnarray}
where $q$ denotes the quark fields with two flavors, $u$ and
$d$, and three colors; $\hat{m}_{0}=\texttt{diag}(m_{u},\, m_{d})$ in flavor space.
The covariant derivative is defined by $D_{\mu}=\partial_{\mu}-i A_{\mu}$
with the background gluon field $A_{\mu}=\delta_{\mu,0}A_0$ supposed constant and uniform.
The temperature--dependent Polyakov effective potential,
$\mathcal{U}(\Phi[A],\bar{\Phi}[A],T) $, is a function of the Polyakov loop $\Phi[A]$ and
its hermitian conjugate $\bar{\Phi}[A]$.
In some analogy with the nonlinear Walecka model, in this study we also try to
include in the NJL term the isoscalar--vector and isovector--vector
interaction channels as given in Ref.~\cite{Ohnishi11}.

Here some considerations are in order.
In our previous work, Refs.~\cite{Shao111,Shao112}, where only the scalar interaction channels were considered,
calculations were performed within a relativistic mean field approximation which
essentially corresponds to the Hartree approximation.
However, the inclusion
of the Fock (exchange) terms would be desirable. In fact, these terms
originate from the correlations due to the Fermi--Dirac statistics, therefore
they are related to a genuine quantum effect, which in general cannot
be neglected when studying a many--body system.
Including the Fock terms, the whole variety of processes in quark
dynamics arising from the fermionic intrinsic degrees of freedom
(~spin, flavor and color~), are automatically accounted for. Indeed, even starting
from an effective Lagrangian containing only scalar channels, the exchange
terms naturally yield contributions in the vector channels, as we will show in the
following.

A quantity of interest in the study of quark dynamics is the statistical
average of the canonical energy-momentum density tensor,
$<: T_{\mu\nu}(x) :>$, from which thermodynamical quantities, such as the
pressure, can be derived.
For the considered Lagrangian, the interaction part of this quantity, $<: T_{\mu\nu}^{I}(x) :>$,
depends on the statistical average of the product of four quark fields, that
in the Hartree-Fock ( HF ) approximation, can be written as:
%\begin{eqnarray}
%<:{\bar q}_\alpha(x)q_\alpha(x){\bar q}_\gamma(x)q_\gamma(x):> =
%<:{\bar q}_\alpha(x)q_\alpha(x):><:{\bar q}_\gamma(x)q_\gamma(x):> \nonumber \\
%- <:{\bar q}_\alpha(x)q_\gamma(x):><:{\bar q}_\gamma(x)q_\alpha(x):>,
%\end{eqnarray}
%
\begin{eqnarray}
&<:{\bar q}_\alpha(x)q_\alpha(x){\bar q}_\gamma(x)q_\gamma(x):>\,\,\,\,\,\,\,\,\,\,\,\,\,
\,\,\,\,\,\,\,\,\,\,\,\,\,\,\,\,\,\,\,\,\,\,\,\,\,\,\,\,\,\,\,\,
\,\,\,\,\,\,\,\,\,\,\,\,\,\,\,\,\,\,\,\,\, \nonumber \\
&=<:{\bar q}_\alpha(x)q_\alpha(x):><:{\bar q}_\gamma(x)q_\gamma(x):>  \nonumber \\
&- <:{\bar q}_\alpha(x)q_\gamma(x):><:{\bar q}_\gamma(x)q_\alpha(x):>,
\end{eqnarray}
where the brackets denote statistical averaging and the colons denote
normal ordering.
It is useful to define the matrix $[{\widehat F}(x)]_{\alpha\beta}$:
$$[{\widehat F}(x)]_{\alpha\beta}=\,
<:\bar{q}_\beta(x)q_\alpha(x):>$$
where $\alpha$ and $\beta$ are triple indices for spin, isospin (flavor)
and color.

As far as one is concerned with the equilibrium properties of isotropic and non--colored
quark matter it is sufficient to consider only the scalar and vector channels
(~isoscalar and isovector~). Then the matrix ${\hat F}(x,p)$
%According to the Clifford algebra,
%it
can be decomposed as:
\begin{eqnarray}\label{clif}
{\hat F}(x)&=& F(x)+{\gamma_\mu}{F^\mu}(x) \nonumber \\
&&+\vec{\tau}\cdot\vec{B}(x)+{\gamma_\mu}\vec{\tau}\cdot\vec{B}^{\mu}(x)
\end{eqnarray}
%$${\hat F}(x)= F(x)+{\gamma_\mu}{F^\mu}(x)
%%+{1\over2}{\sigma_{\mu\nu}}F^{\mu\nu}(x)
%$$
%\begin{equation}
%+\vec{\tau}\cdot\vec{B}(x)+{\gamma_\mu}\vec{\tau}\cdot\vec{B}^{\mu}(x)~.
%%+\gamma^5A(x)-{\gamma^5\gamma_\mu}{A^\mu}(x)~.
%\label{clif}
%\end{equation}
%It should be noticed that this function is related to the various densities
%with their specific transformation
%properties both in ordinary space and in isospin space (~isoscalar
%and isovector~).
%For instance
It should be noticed that this matrix is related to the various densities
characterizing the system. Indeed
the scalar and current isoscalar densities are given by:
$$\rho_S(x)=\,<:{\bar q(x)}q(x):>=\,Tr{\widehat F}(x) = 8\,N_c\,F(x)\, ,$$
$$j^{\mu}(x)=\,<:{\bar q(x)}\gamma^{\mu}q(x):>=
\,Tr\gamma^{\mu}{\widehat F}(x) = 8\,N_c\,F^{\mu}(x)\, ,$$
while the isovector counterparts are given by:
$$\rho_3(x)=\,<:{\bar q(x)}\tau_3q(x):>=\,Tr\tau_3{\widehat F}(x)
= 8\,N_c\,B_3(x)\, ,$$
$$j^{\mu}_3(x)=\,<:{\bar q(x)}\tau_3\gamma^{\mu}q(x):>=
\,Tr\gamma^{\mu}\tau_3{\widehat F}(x) = 8\,N_c\,B_3^{\mu}(x)\, ,$$
where the traces are taken over spin, flavor and color indices.
Then the interaction part of
the energy--momentum density tensor, in the HF approximation, reads:
%
%\begin{eqnarray}
%<: T_{\mu\nu}^{(I)}(x) :> =
% T_{\mu\nu}^{(I)}(x)_{Hartree}\nonumber \\
%%8\int d^4 p~ p_{\nu}F_{\mu}(x,p) +
%%+ {f_S\over 2}
%%\rho_S^2 g_{\mu\nu}
%%- {f_V\over 2}
%%j_\lambda(x)
%%j^\lambda(x) g_{\mu\nu} \nonumber \\
%%%\end{equation}
%%%aa
%- [2G_\sigma Tr\big({\widehat F}(x)
%{\widehat F}(x) +(i\gamma_5)\vec{\tau}{\widehat F(x)}
%\cdot(i\gamma_5) \vec{\tau}{\widehat F}(x)\big)\nonumber \\
%- 2G_\omega Tr\big(\gamma_\lambda{\widehat F(x)}
%\gamma^\lambda {\widehat F}(x) +
%\gamma_5\gamma_\lambda{\widehat F(x)}
%\gamma_5\gamma^\lambda {\widehat F}(x)\big)\nonumber \\
%+2G_\delta Tr\big({\vec{\tau}\widehat F}(x)\cdot
%\vec{\tau}{\widehat F}(x) + (i\gamma_5){\widehat F(x)}
%(i\gamma_5) {\widehat F}(x)\big)\nonumber \\
%- 2G_\rho Tr\big(\vec{\tau}\gamma_\lambda{\widehat F(x)}
%\cdot\vec{\tau}\gamma^\lambda {\widehat F}(x) +
%\vec{\tau}\gamma_5\gamma_\lambda{\widehat F(x)}
%\cdot\vec{\tau}\gamma_5\gamma^\lambda {\widehat F}(x)\big)
%]g_{\mu\nu}
%%+ f_\delta ... - f_\omega ...
%\end{eqnarray}
%
%
%
\begin{widetext}
\begin{eqnarray}
<: T_{\mu\nu}^{(I)}(x) :> &=&
 T_{\mu\nu}^{(I)}(x)_{Hartree}
%8\int d^4 p~ p_{\nu}F_{\mu}(x,p) +
%+ {f_S\over 2}
%\rho_S^2 g_{\mu\nu}
%- {f_V\over 2}
%j_\lambda(x)
%j^\lambda(x) g_{\mu\nu} \nonumber \\
%%\end{equation}
%%aa
- \bigg[2G_\sigma Tr\big({\widehat F}(x)
{\widehat F}(x) +(i\gamma_5)\vec{\tau}{\widehat F(x)}
\cdot(i\gamma_5) \vec{\tau}{\widehat F}(x)\big)\nonumber \\
&&- 2G_\omega Tr\big(\gamma_\lambda{\widehat F(x)}
\gamma^\lambda {\widehat F}(x) +
\gamma_5\gamma_\lambda{\widehat F(x)}
\gamma_5\gamma^\lambda {\widehat F}(x)\big)
+2G_\delta Tr\big({\vec{\tau}\widehat F}(x)\cdot
\vec{\tau}{\widehat F}(x) + (i\gamma_5){\widehat F(x)}
(i\gamma_5) {\widehat F}(x)\big)\nonumber \\
&&- 2G_\rho Tr\big(\vec{\tau}\gamma_\lambda{\widehat F(x)}
\cdot\vec{\tau}\gamma^\lambda {\widehat F}(x) +
\vec{\tau}\gamma_5\gamma_\lambda{\widehat F(x)}
\cdot\vec{\tau}\gamma_5\gamma^\lambda {\widehat F}(x)\big)
\bigg]g_{\mu\nu}.
%+ f_\delta ... - f_\omega ...
\end{eqnarray}
\end{widetext}

The exchange terms that appear in the energy--momentum tensor can be
evaluated exploiting the decomposition given in Eq.~(\ref{clif}).
After some algebra one realizes that the effect of the Fock terms is equivalent
to redefine the coupling constants, as written below:
\begin{eqnarray}
{\tilde G_\sigma} = G_\sigma + (G_\sigma-G_\delta)/12\nonumber \\
{\tilde G_\delta} = G_\delta - (G_\sigma-G_\delta)/12\nonumber \\
{\tilde G_\omega} = G_\omega(1+1/6) + (G_\sigma+G_\delta)/6 + G_\rho/2\nonumber \\
{\tilde G_\rho} = G_\rho(1-1/6) + G_\omega/6~.
\end{eqnarray}
Then, with the effective coupling constants given above, calculations can be performed
as in the Hartree approximation.

It should be remarked that all the relevant interaction channels in
general can occur in the HF approximation, even if some channel is
absent in the original
Lagrangian. For instance, if the vector channels are not present,
i.e., we take $G_\omega = 0$ and $G_\rho = 0$ as in our previous work,
contributions to the $\omega$ channel are naturally arising from the exchange
terms associated with the scalar channels.
On the other hand, the vector isovector $\rho$ channel gets contributions
(both direct and exchange) only from vector channels.

Since the critical end--point of the first order chiral transition
appreciably depends on the strength of
the vector channel interaction \cite{Kita02,Sasa07,Fuku08},
the exchange contribution $(G_\sigma+G_\delta)/6$ to the effective value
of ${\tilde G_\omega}$ could represent a reference value
for the vector channel interaction \cite{fock}.

In the following we will adopt the choice ${\tilde G_\delta} = 0$ and the notation
$G\equiv{\tilde G_\sigma} = (G_\sigma+G_\delta)$.
The coupling constants of the vector interactions,
${\tilde G_\omega}$ and ${\tilde G_\rho}$,
will be taken as parameters, and different
values will be used to investigate their influence on the phase transition.
For convenience we define $r_\omega={\tilde G_\omega}/G,\,r_\rho={\tilde
G_\rho}/G$.

For the temperature dependent effective potential
$\mathcal{U}(\Phi,\bar{\Phi},T)$ we use the parametrization given in Ref.~\cite{Robner07}

\begin{eqnarray}
     \frac{\mathcal{U}(\Phi,\bar{\Phi},T)}{T^4}&=&-\frac{a(T)}{2}\bar{\Phi}\Phi
                                                +b(T)\mathrm{ln}[1-6\bar{\Phi}\Phi\nonumber\\
                                                &&+4(\bar{\Phi}^3+\Phi^3)-3(\bar{\Phi}\Phi)^2],
\end{eqnarray}
where
\begin{equation}
    a(T)=a_0+a_1\bigg(\frac{T_0}{T}\bigg)+a_2\bigg(\frac{T_0}{T}\bigg)^2,\ \  b(T)=b_3\bigg(\frac{T_0}{T}\bigg)^3.
\end{equation}
%and
%\begin{equation}
%    b(T)=b_3\bigg(\frac{T_0}{T}\bigg)^3.
%\end{equation}
The parameters $a_i$, $b_i$ are fitted to the lattice QCD results in pure gauge theory
at finite temperature. In the equation above $T_0$ represents the temperature where
the Polyakov potential gives a deconfinement phase transition in a pure gauge theory.
The original value of $T_0$ fitted to pure gauge lattice QCD data is $270 MeV$
~\cite{Fukugita90}. When
fermion fields are included, the temperature $T_0$ is usually rescaled to obtain
a consistent result with the full lattice data, which give the value $T^c=173\pm 8 MeV$
for deconfiniment transition temperature \cite{Karsch01,Karsch02,Kaczmarek05}.
In this paper the value $210 MeV$ for $T_0$ is adopted.

The PNJL model is not renormalizeable, so a cut-off $\Lambda$ is introduced to get
finite results for three--momentum space integrations. For the model parameters
we take the values $\Lambda=651 MeV$, $G=5.04\mathrm{GeV}^{-2}$,
$m_{u,d}=5.5 MeV$, determined by fitting the chiral condensate, $f_{\pi}$ 
and $M_{\pi}$ to their experimental values~\cite{Ratti06}.
The coefficients in the Polyakov effective potential are listed 
in Table \ref{tab:1}.
\begin{table}[ht]
\tabcolsep 0pt \caption{\label{tab:1}Parameters in Polyakov effective potential given in~\cite{Robner07}}
\setlength{\tabcolsep}{2.5pt}
\begin{center}
\def\temptablewidth{0.58\textwidth}
%{\rule{\temptablewidth}{0.5pt}}
\begin{tabular}{c c c c}
\hline
\hline
   {$a_0$}                      & $a_1$        & $a_2$      & $a_3$           \\  \hline
   $ 3.51$                   & -2.47        &  15.2      & -1.75               \\ \hline
\hline
\end{tabular}
 % {\rule{\temptablewidth}{0.5pt}}
\end{center}
\end{table}

The thermodynamical--potential density of quark matter in the mean field
approximation reads
\begin{widetext}
\begin{eqnarray}
\Omega&=&\mathcal{U}(\bar{\Phi}, \Phi, T)+G({\phi_{u}+\phi_{d}})^{2}-
{\tilde G}_\omega(\rho_{u}+\rho_{d})^{2}
-{\tilde G}_\rho(\rho_{u}-\rho_{d})^{2}-2\int_\Lambda \frac{\mathrm{d}^{3}p}{(2\pi)^{3}}3(E_u+E_d) \nonumber \\
&&-2T \sum_{u,d}\int \frac{\mathrm{d}^{3}p}{(2\pi)^{3}} \bigg[\mathrm{ln}(1+3\Phi e^{-(E_i-\mu_i^*)/T}+3\bar{\Phi} e^{-2(E_i-\mu_i^*)/T}+e^{-3(E_i-\mu_i^*)/T}) \bigg]\nonumber \\
&&-2T \sum_{u,d}\int \frac{\mathrm{d}^{3}p}{(2\pi)^{3}} \bigg[\mathrm{ln}(1+3\bar{\Phi} e^{-(E_i+\mu_i^*)/T}+3\Phi e^{-2(E_i+\mu_i^*)/T}+e^{-3(E_i+\mu_i^*)/T}) \bigg],
\end{eqnarray}
\end{widetext}
where $\rho_i$ is the number density of quarks of flavor $i$,
$E_i=\sqrt{\vec{p}^{\,2}+M_i^2}$ and $\mu_i^*$ are the corresponding
energy--momentum dispersion relation and effective chemical potential
with
\begin{equation}\label{muu}
\mu_u^*=\mu_u-2{\tilde G}_\omega(\rho_u+\rho_d)-2{\tilde G}_\rho(\rho_u-\rho_d)
\end{equation}
\begin{equation}\label{mud}
\mu_d^*=\mu_d-2{\tilde G}_\omega(\rho_u+\rho_d)+2{\tilde G}_\rho(\rho_u-\rho_d)
\end{equation}
The dynamical quark masses and quark condensates are
coupled with the following equations\newline
\begin{equation}
M=m_{0}-2G(\phi_u+\phi_d),
\label{mass}
\end{equation}
\begin{equation}
\phi_i=-2N_{c}\int\frac{d^{3}\boldsymbol k}{(2\pi)^{3}}\frac{M}{E}
\big(1-n_i(k)-\bar{n}_i(k)\big),
\end{equation}
where $n_i(k)$ and $\bar{n}_i(k)$ are modified Fermion distribution
functions of quark and antiquark (similar to those given in Ref.~\cite{Hansen07}
without the vector contributions)
\begin{widetext}
\begin{equation}
  n_{i}(k)=\frac{\Phi e^{-(E_i-\mu_i^*)/T}+2\bar{\Phi} e^{-2(E_i-\mu_i^*)/T}+e^{-3(E_i-\mu_i^*)/T}}{1+3\Phi e^{-(E_i-\mu_i^*)/T}+3\bar{\Phi} e^{-2(E_i-\mu_i^*)/T}+e^{-3(E_i-\mu_i^*)/T}} ,
\end{equation}
\begin{equation}
  \bar{n}_{i}(k)=\frac{\bar{\Phi} e^{-(E_i+\mu_i^*)/T}+2{\Phi} e^{-2(E_i+\mu_i^*)/T}+e^{-3(E_i+\mu_i^*)/T}}{1+3\bar{\Phi} e^{-(E_i+\mu_i^*)/T}+3{\Phi} e^{-2(E_i+\mu_i^*)/T}+e^{-3(E_i+\mu_i^*)/T}} ,
\end{equation}
\end{widetext}
The values of $\phi_u, \phi_d, \Phi$ and $\bar{\Phi}$ are determined
by minimizing the thermodynamical potential
\begin{equation}
\frac{\partial\Omega}{\partial\phi_u}=\frac{\partial\Omega}{\partial\phi_d}=\frac{\partial\Omega}{\partial\Phi}
=\frac{\partial\Omega}{\partial\bar\Phi}=0.
\end{equation}
All the thermodynamic quantities relevant to the bulk properties of
quark matter can be obtained from $\Omega$. Particularly, we note that
the pressure and energy density should be zero in the vacuum.

The number density of quarks of flavor $i$
\begin{equation}
\rho_{i}=2\times3\int\frac{d^{3}\boldsymbol k}{(2\pi)^{3}}(n_{i}(k)-
\bar{n}_{i}(k))
\label{quarkdensity}
\end{equation}
can be derived by means of the relation
$\rho_{i}=-\partial \Omega_{}/\partial \mu_{i}$.
The baryon and isospin densities and the corresponding chemical potentials
in quark phase are defined by
\begin{equation}
\rho_{B}^Q=\frac{1}{3}(\rho_u+\rho_d),\ \ \ \ \rho_{3}^Q=
\rho_u-\rho_d,
\end{equation}
\begin{equation}
\mu_{B}^Q=\frac{3}{2}(\mu_u+\mu_d),\ \ \ \ \mu_{3}^Q=\frac{1}{2}(\mu_u-\mu_d)\, ,
\end{equation}
while the asymmetry parameter of quark matter is defined by
\begin{equation}
   \alpha^{Q}\equiv-\frac{\rho_{3}^{Q}}{\rho_{B}^{Q}}=3\frac{\rho_d-\rho_u}
{\rho_u+\rho_d}.
\end{equation}
Analogous definitions hold for hadronic matter:
\begin{equation}
\rho_{B}^H=\rho_p+\rho_n,\ \ \ \ \rho_{3}^H=
\rho_p-\rho_n,
\end{equation}
\begin{equation}
\mu_{B}^H=\frac{1}{2}(\mu_p+\mu_n),\ \ \ \
\mu_{3}^H=\frac{1}{2}(\mu_p-\mu_n)\, ,
\end{equation}
\begin{equation}
   \alpha^{H}\equiv-\frac{\rho_{3}^{H}}{\rho_{B}^{H}}=\frac{\rho_n-\rho_p}
{\rho_n+\rho_p}\, ,
\end{equation}
where $\rho_p$ and $\rho_n$ are the proton and the neutron densities,
respectively.
%The above introduction is a separate description of the hadronic and quark
%phase.
When a mixed phase of quarks and hadrons is considered,
the Gibbs' conditions (thermal, chemical and mechanical equilibrium)
\begin{eqnarray}\label{tcm}
& &\mu_B^H(\rho_B^{},\rho_3^{},T)=\mu_B^Q(\rho_B^{},\rho_3^{},T)\nonumber\\
& &\mu_3^H(\rho_B^{},\rho_3^{},T)=\mu_3^Q(\rho_B^{},\rho_3^{},T)\nonumber\\
& &P^H(\rho_B^{},\rho_3^{},T)=P^Q(\rho_B^{},\rho_3^{},T),
\end{eqnarray}
should be fulfilled (a general discussion
of phase transitions in multicomponent systems can be found
in Ref.~\cite{Glendenning92}). In Eq.~(\ref{tcm}),
$\rho_B^{}=(1-\chi)\rho_B^{H}+\chi \rho_B^{Q}$ is the total
baryon density and  $\rho_3^{}=(1-\chi)\rho_3^{H}+\chi \rho_3^{Q}$
is the total isospin density, where $\chi$ is the quark fraction.
In heavy-ion collisions, for a given initial charge asymmetry
the global asymmetry parameter $\alpha$ of the mixed phase
\begin{eqnarray}
& &   \alpha\equiv-\frac{\rho_{3}^{}}{\rho_{B}^{}}= \frac{(1-\chi)\rho_3^{H}+
\chi \rho_3^{Q}}{(1-\chi)\rho_B^{H}+\chi \rho_B^{Q}},
\end{eqnarray}
should be constant according to the charge conservation,
%of isospin,
but the asymmetry parameters $\alpha^H,\alpha^Q$ in the
separate phases can vary with $\chi$. For details, one can
refer to Refs.~\cite{Torohq11,Shao111,Shao112}.

\section{NUMERICAL RESULTS AND DISCUSSIONS}

In this section, we display the numerical results and
discuss the influence of isovector and isoscalar vector interactions
on the phase diagram of the hadron-quark phase transition.
%In the calculation, the asymmetric parameter $\alpha=0.2$ is
%taken for asymmetric matter, and the largest asymmetry
%$\alpha=0.227$ is possibly reached in heavy-ion collision experiment
%for $^{238}\mathrm{U}+^{238}\mathrm{U}$ collision.

In actual calculations a value of $0.2$ is chosen for the asymmetry parameter
$\alpha$. We notice that in heavy--ion collision experiments the largest
value of $\alpha$, $\alpha=0.227$, is possibly reached in
$^{238}\mathrm{U}+^{238}\mathrm{U}$ collisions.

\subsection{The role of the isovector vector interaction}

Firstly, we only focus on the influence of the isovector vector
interaction on the phase
transition in the two-phase model, then we  set
$r_\omega=0$, and perform calculations for various values of $r_\rho$,
$r_\rho=0.0,\,0.25,\,0.5,\,1.0$.
\begin{figure}[htbp]
\begin{center}
\includegraphics[scale=0.3]{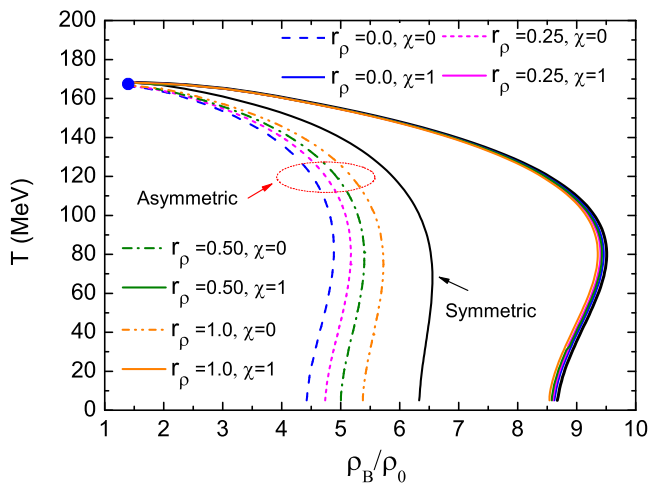}
\caption{\label{fig:V0RHO-T}{(Color online) Diagram of the hadron--quark phase
transition in the $T-\rho_B^{}$ plane
for symmetric~($\alpha=0$) and asymmetric matter ($\alpha=0.2$)
with $r_\rho=0.0,\,0.25,\,0.5,$ and 1.0. The baryon density $\rho_B^{}$
is expressed in units of the density of ordinary nuclear matter $\rho_0$.
The lines in the left side corresponding to
$\chi=0$ represent the onset of the mixed phase, and those
in the right side corresponding to $\chi=1$ denote the beginning of
the pure quark phase. The dot indicates the critical end point.}}
\end{center}
\end{figure}

We display the phase diagram in $T-\rho_B^{}$ plane in
Fig.~\ref{fig:V0RHO-T}. In agreement with the results of
Refs.~\cite{Shao111,Shao112},
this figure shows that the onset density of the mixed phase in asymmetric
matter  is smaller than that of symmetric matter. This means that
in heavy--ion collision experiments the phase transition could be
relatively easier to be reached for asymmetric matter with respect to
the symmetric case. However, the onset density appreciably depends on the
strength of the interaction in the isovector--vector channel. It is
shifted to higher densities with increasing the coupling parameter
${\tilde G}_\rho$. Meanwhile the density range of the mixed phase
shrinks. Indeed, similarly to what is observed in the hadron sector,
the isovector vector channel yields a positive
(repulsive) contribution to the pressure of the quark phase, connected
to the corresponding symmetry energy.
\begin{figure}[htbp]
\begin{center}
\includegraphics[scale=0.3]{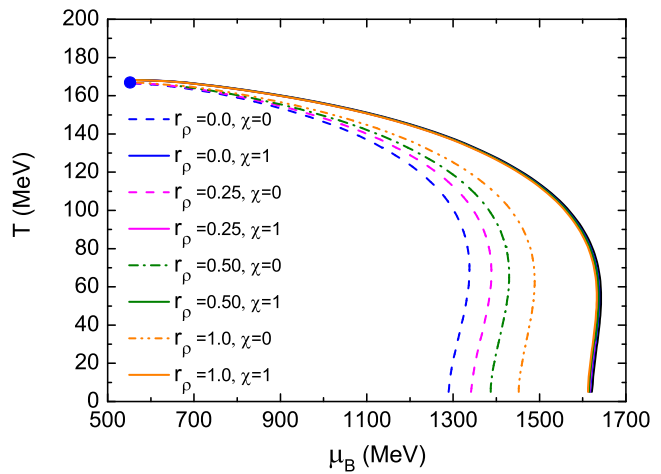}
\caption{\label{fig:V0MU-T}{(Color online) The same of Fig. 1 but for the
$T-\mu_B^{}$ plane.}}
\end{center}
\end{figure}

Similar features are observed for the phase diagram in the $T-\mu_B^{}$ plane
as shown in Fig.~\ref{fig:V0MU-T}. However, we notice that
for symmetric matter, only one phase--transition line exists. This is due to
the fact that in the symmetric case the curve $T-\mu_B$ does not depend on
the quark fraction $\chi$, see also Refs.~\cite{Shao111,Shao112}.
%**(Comment: the following statement is not clear to me)**
%but two for asymmetric matter
%because there are two conservation charges, baryon
%and isospin conservations.
Both figures 1 and 2 show that the isovector--vector channel of the
quark interaction plays a more important role on the
phase transition with increasing density and lowering temperature. In fact,
at higher temperatures and/or lower densities the role of interactions
in general becomes weaker with respect to the kinetic contributions. Finally,
we still observe the occurrence of a critical end point (CEP) of the first order
phase transition, analogously to what happens when only scalar channels are
considered \cite{Shao111,Shao112}. 
%Including the isovector--vector
%channel in the interaction the CEP is slightly shifted to the left.

\begin{figure}[htbp]
\begin{center}
\includegraphics[scale=0.3]{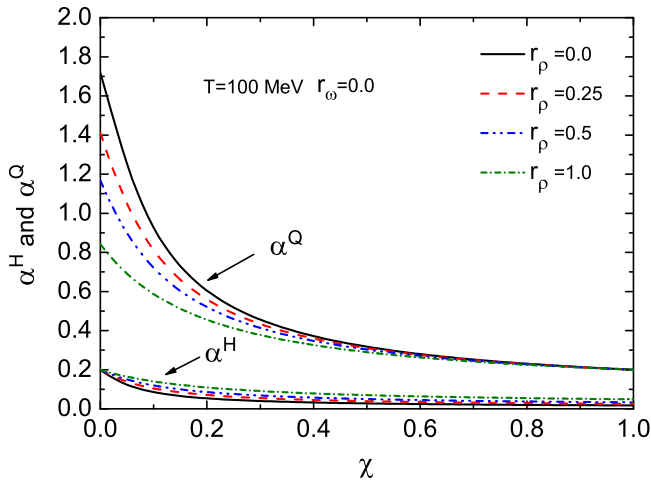}
\caption{\label{fig:V0chi-alphaQ}{(Color online) 
Asymmetry of hadronic and quark
matter in the mixed phase as a function of the quark concentration for
various values of the coupling constant in the isovector--vector interaction
channel.}}
\end{center}
\end{figure}

In Fig.~\ref{fig:V0chi-alphaQ} the asymmetry parameters of hadronic and
quark matter in the mixed phase are displayed as a function of the quark
fraction $\chi$, for the temperature $T=100~MeV$.  We can observe a clear
Isospin Distillation effect~\cite{Toro06,Torohq11}, i.e. the
asymmetry of quark matter is much larger than 0.2 at the beginning
of the phase transition and decreases with increasing the quark
fraction. Whereas the asymmetry of the hadronic matter keeps below 0.2
and is a slowly decreasing 
%decreases slowly as a
function of $\chi$. These features of  the local asymmetry may lead to
some observable effects in the hadronization during the expansion phase of
heavy ion collisions,
such as an inversion in the trend of
emission of neutron rich clusters, an enhancement of $\pi^-/\pi^+$,
$K^0/K^+$ yield ratios in high-density regions, as well as an
enhancement of the production of isospin-rich resonances and
subsequent decays, for more details see
Refs.~\cite{Torohq11,Shao111,Shao112}.
These signals are possible to be probed in the
newly planned facilities, such as FAIR at GSI-Darmstadt and NICA at
JINR-Dubna.

However we also see that the strengthening of the coupling
constant ${\tilde G}_\rho$ reduces the distillation effect and
consequently the asymmetry parameter $\alpha^Q$, which
may weaken the observational signals of the phase transition,
although always present. The uncertainty is that the relevant
$\rho$-coupling constant cannot be unambiguously determined 
from lattice--QCD calculations.
%in theory.
So the observation or not of the related signals in
experiments can provide some hints on this aspect.

In any case we note that both isospin effects, earlier density
transition and isospin distillation are still there even with large
values of $r_\rho$. They appear as general, quite robust  effects of 
asymmetric matter,
not much affected by the introduction of a symmetry 
interaction term in the quark sector. In order to confirm
all that, in the next paragraph we present other results obtained in
a much simpler Iso-MIT-Bag model.

%\subsubsection*
\subsection{Few results with the MIT-Bag model}
As already remarked the isospin effects on the mixed phase
(boundaries and asymmetries in the two phases) are naturally related
to the presence of a symmetry repulsive term in the quark sector. In
order to confirm that this is a general result, not depending on the
different quark models, we present here few similar calculations of
the hadron-quark transition, in a two-EoS approach, using the
MIT-Bag model for the quark matter
\cite{Toro06,Toro09,Torohq11,Liu11}.

Now the results are very sensitive to the choice of the Bag-constant
$B$, in particular at low temperatures and high baryon densities. In
fact for low B-values we can even get a disappearing of the
transition since the hadron pressure cannot match anymore the quark
pressure, as discussed in detail in Ref.~\cite{Liu11}.

Here we choose a Bag-constant $B=(160MeV)^4$, which gives for
symmetric matter a high-density mixed phase structure very close to
the one of Fig.~\ref{fig:V0RHO-T}, obtained with the $PNJL$ model.
Of course the same $NL\rho\delta$ Relativistic Mean Field
interaction is used for the hadron sector.

A quark isovector-vector ($\rho$-like) term is introduced by a naive
application of a constituent quark model of the nucleons, i.e., just
reducing of a factor 3 the Nucleon-$\rho$  coupling constant of the
hadron part. As a consequence we get straightforward corrections to
the quark pressure and chemical potentials with respect to the
simple relativistic Fermi gas values of the MIT-Bag model ($P^Q(F),
\mu_u(F), \mu_d(F)$):
\begin{eqnarray}\label{ISOE}
& & P^Q = P^Q(F) + \frac{g_{\rho,q}^2}{2m_\rho^2} (\rho_u-\rho_d)^2, \nonumber\\
& &\mu_u = \mu_u(F) + \frac{g_{\rho,q}^2}{m_\rho^2} (\rho_u-\rho_d),  \nonumber\\
& &\mu_d = \mu_d(F) - \frac{g_{\rho,q}^2}{m_\rho^2} (\rho_u-\rho_d),
\end{eqnarray}
with $g_{\rho,q}=g_{\rho,N}/3$ and $m_\rho$ is the $\rho$-meson
mass. The coupling choice is fixed by the $NL\rho\delta$
parametrization of the hadron sector,
$$
f_{\rho,N} = \frac{g_{\rho,N}^2}{m_\rho^2} = 3.15 fm^2,
$$
see the detailed Appendix A of Ref.~\cite{Liu11}. We note that such
simple insertion of isovector-vector terms in quark phase
(Iso-MIT-Bag model, in the following named $ISOE$ results) can be
particularly justified at high baryon densities and chemical
potentials, where we expect a more relevant role of the hadronic
degrees of freedom. In any case here our aim is to show general
quark symmetry energy effects on the hadron-quark transition at low
temperatures in isospin asymmetric matter, that confirm the results
of the previous section with the Hadron-PNJL approach.

\begin{figure}[htbp]
\begin{center}
\includegraphics[scale=0.3]{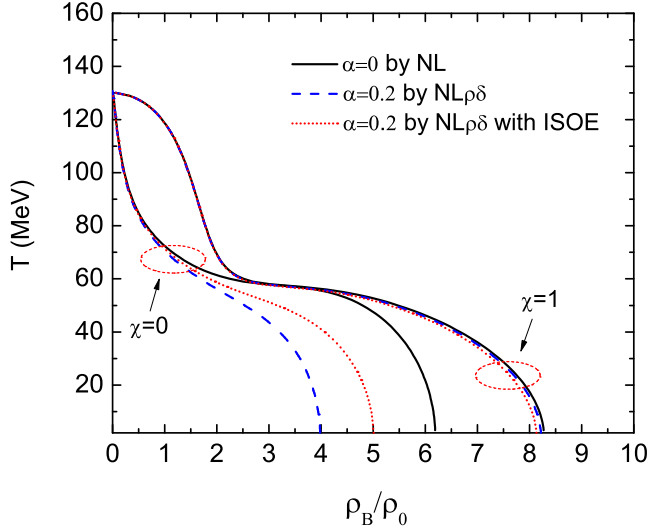}
\caption{\label{fig:Trho-isoMIT} (Color online) $NL\rho\delta$-MITBag results:
$T-\rho_B^{}$ plane of the hadron-quark phase transition for
symmetric~($\alpha=0$) and asymmetric matter ($\alpha=0.2$) with the
ISOE choice for the $\rho$-coupling. The curves in the left side
with $\chi=0$ represent the beginning of the mixed phase, and those
in the right side with $\chi=1$ mean the beginning of the pure quark
phase.}
\end{center}
\end{figure}

The Fig.~\ref{fig:Trho-isoMIT} is the $T - \rho_B^{}$ phase diagram for
symmetric and $\alpha=0.2$ isospin asymmetric matter,
corresponding to the previous Fig.~\ref{fig:V0RHO-T} obtained with
the PNJL model in the quark sector, with various weights of the
$\rho$ coupling. We clearly see that at temperatures below 40 MeV
and at high baryon densities the curves are very similar, in
particular in the choice in Fig.~\ref{fig:V0RHO-T} of a $r_\rho$
ratio equal to 0.5, close to the evaluation around 0.7 we use in the
MIT-Bag model following the constituent quark picture.

\begin{figure}[htbp]
\begin{center}
\includegraphics[scale=0.3]{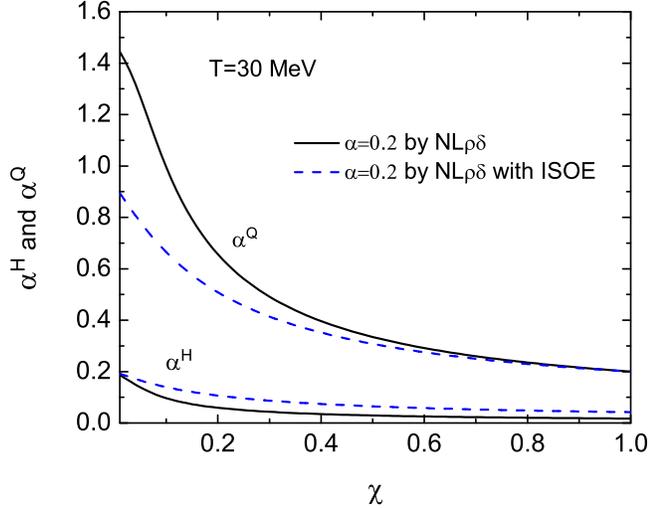}
\caption{\label{fig:alphachi-isoMIT}(Color online) 
$NL\rho\delta$-MITBag results:
Asymmetry of hadronic and quark matter in the mixed phase as
functions of quark concentration without and with the ISOE choice for the
isovector vector interaction coupling.}
\end{center}
\end{figure}

The same comment is valid for the behavior of the asymmetry
parameters in the mixed phase, shown in the ISOE calculation in the
Fig.~\ref{fig:alphachi-isoMIT}, to compare to the previous
Fig.~\ref{fig:V0chi-alphaQ}.

In general we can say that all the isospin influence on the
hadron-quark transition is still present when we introduce
vector-isovector terms in the quark phase, even with relatively
large weights. Of course the interaction symmetry repulsion in the
quark sector will reduce both effects, earlier transition densities
and isospin distillation in the mixed phase, but the possibility of
related observations appears still there.

\subsection{The role of the isoscalar vector interaction}

We discuss now the results obtained when the
isoscalar--vector interaction channel in the quark sector is turned on.
We choose the value of $0.2$ for the ratio $r_\omega={\tilde G}_\omega/G$.
This value is close to the contribution to this channel from the exchange
terms of the scalar channels, see Eqs.~(5).
\begin{figure}[htbp]
\begin{center}
\includegraphics[scale=0.3]{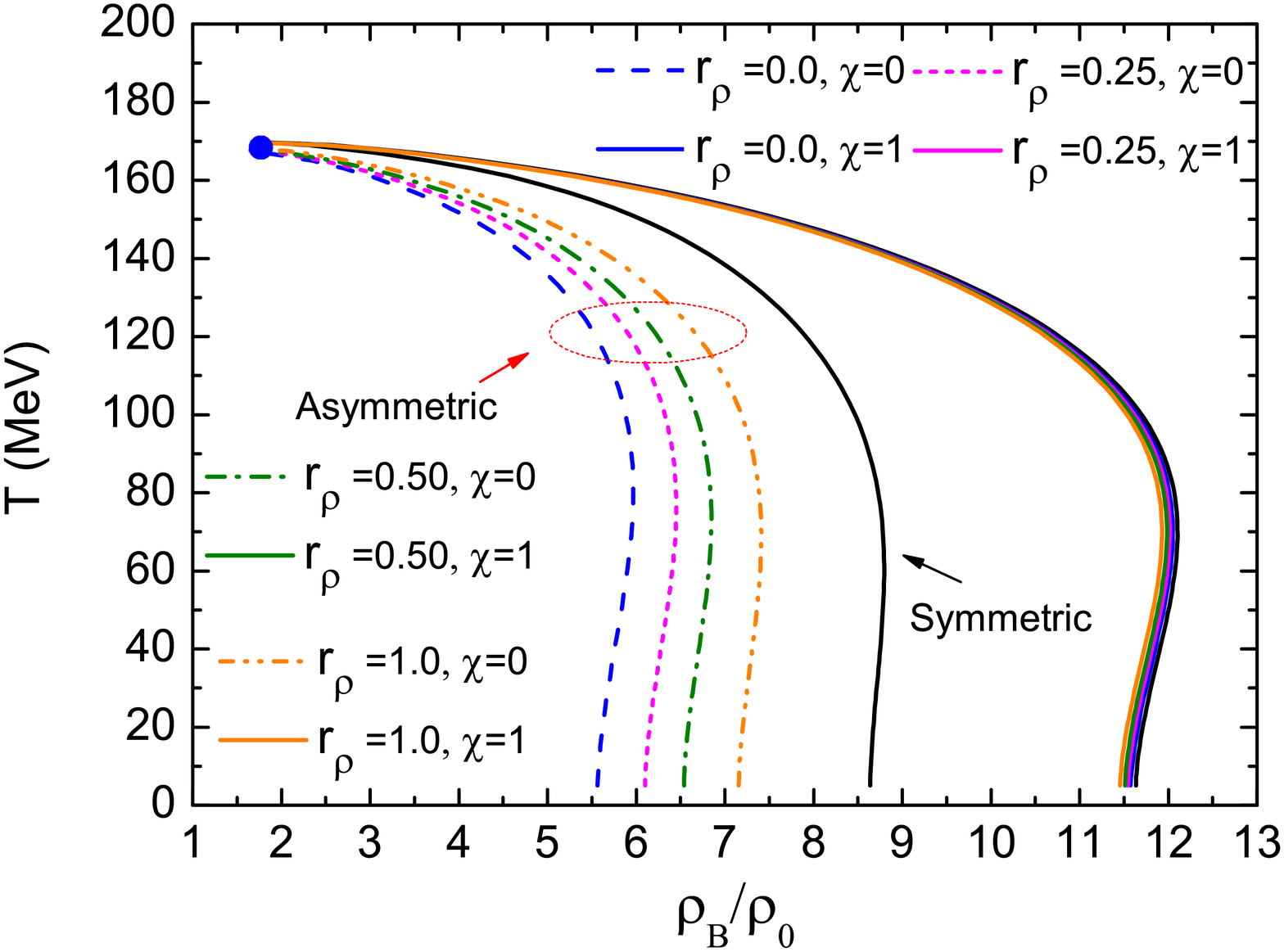}
\caption{\label{fig:V2RHO-T} (Color online) The same as Fig.~\ref{fig:V0RHO-T}
except for the inclusion of isoscalar vector interaction with $r_\omega=0.2$}
\end{center}
\end{figure}

\begin{figure}[htbp]
\begin{center}
\includegraphics[scale=0.3]{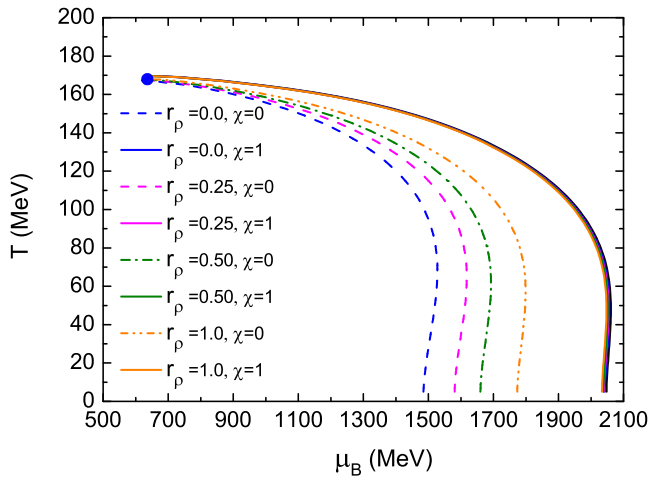}
\caption{\label{fig:V2MU-T} (Color online) The same as Fig.~\ref{fig:V0MU-T}
except for the inclusion of isoscalar vector interaction with $r_\omega=0.2$}
\end{center}
\end{figure}

In Figs.~\ref{fig:V2RHO-T}
and \ref{fig:V2MU-T} are displayed the diagrams of the hadron--quark phase
transition in the $T-\rho_B$ and $T-\mu_B$ planes respectively. Compared to
Figs.~\ref{fig:V0RHO-T} and \ref{fig:V0MU-T} the phase--transition curves
are significantly moved toward higher values of density/chemical potential.
This can be explained in terms of the repulsive
contribution of the isoscalar--vector channel to the quark energy  and, as a consequence, to the chemical
potential (~see Eqs.~\ref{muu} and \ref{mud}~). More specifically, the
relevant quantity in the kinetic contribution to the thermodinamical
potential, besides the temperature, is the effective chemical potential,
$\mu^*_i$. This quantity is appreciably smaller than the full chemical
potential. Then, higher values of the latter quantity are necessary to fulfil
the Gibbs conditions. Moreover, we obeserve that for low values of temperature
the quark chemical potential attains values of the same order of magnitude
as the cut--off $\Lambda$. However, this does not give rise to inconsistency,
since the momentum scale is set by the Fermi momentum, which is determined
by the effective chemical potential.
Also the CEP moves to higher values of the density/chemical potential
although slightly, whereas  the corresponding
critical temperature almost keeps the same value. Moreover, in asymmetric matter
the region of coexistence of the two phases is more extended when the isoscalar--vector
interaction channel is included in the quark sector.

\begin{figure}[htbp]
\begin{center}
\includegraphics[scale=0.3]{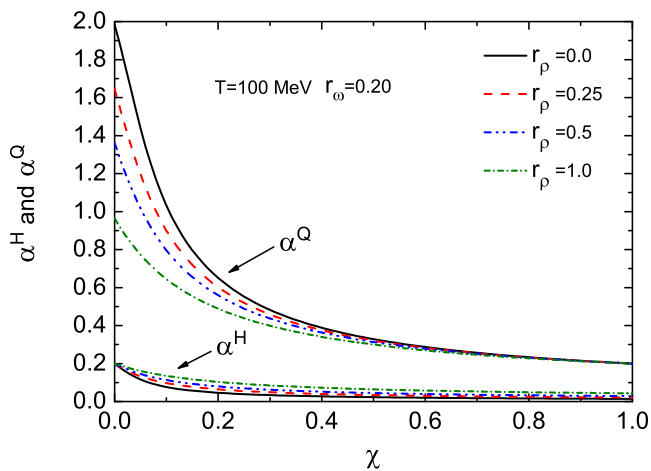}
\caption{\label{fig:V2chi-alphaQ}(Color online) 
The same as Fig.~\ref{fig:V0chi-alphaQ}
except for the inclusion of isoscalar vector interaction with $r_\omega=0.2$}
\end{center}
\end{figure}

Finally, comparing  Fig.~\ref{fig:V2chi-alphaQ} with
Fig.~\ref{fig:V0chi-alphaQ} one can observe that the inclusion of the
isoscalar--vector channel leads to a small increase of the asymmetry parameter
of the quark matter in the mixed phase. The Isospin Distillation effect
may be strengthened by this interaction channel, but the onset of the
mixed phase is shifted toward higher densities.

From the results discussed above it is clear that the inclusion of
the repulsive vector interaction in the quark sector affects significantly
the possibility to observe the first order hadron-quark phase transition,
especially in the low temperature--finite chemical potential region.
Indeed, with $r_\omega= 0.2$, the onset transition density, even in asymmetric matter,
is already shifted to density values larger than $5.5~\rho_0$.
Thus the study of this transition and of the associated phase diagram may
give information on the properties of the interaction between quarks
and corresponding coupling constants.  From the definition of the
effective coupling constants, Eqs.(5), one can see that a considerable
positive contribution to ${\tilde G}_\omega$, equal to $0.17~G$, already
comes from the exchange terms of the scalar channels.
%Since for physical
%reasons (positive symmetry energy) ${\tilde G}_\rho$ must be positive,
On the other hand,
the observation  of such a transition in relativistic heavy ion
collisions at intermediate energies, where moderate density and temperature values
are reached, would point to smaller value of $r_\omega$. Thus such an evidence
could be taken as an indication of
%that is predicted only if $r_\omega$ keeps relatively small, may indicate
the presence of the vector channels already in the original quark Lagrangian,
with a negative $G_\omega$ and a positive $G_\rho$. We note that the sign of $G_\rho$
is dictated by physical reasons: it must be positive in order
to obtain a positive symmetry energy.

\section{Summary}
We have studied the hadron-quark/gluon phase
transition in the two-phase model with the newly added
isovector and isoscalar vector interactions for quark matter.
We stress the presence of vector terms in the quark Equation 
of State just due to exchange contributions from scalar fields
in the effective PNJL Lagrangian.

The consideration of the isovector vector interaction splits the
effective quark chemical potential of $u$ and $d$ quarks. The
calculations show that the phase transition densities are
delayed to higher values and the ranges of the mixed phases
shrink with the increase of the coupling constant ${\tilde G}_\rho$. Meanwhile
the asymmetry parameter $\alpha^Q$ at small $\chi$ is reduced
for a larger ${\tilde G}_\rho$.

Furthermore, with the inclusion of the isoscalar vector interaction,
the whole phase diagrams move to higher baryon densities/chemical
potentials, but the temperatures of the CEP almost keep unchanged.

In our previous study, we have proposed some possibly observable
signals of the hadron-quark/gluon phase transition for asymmetric
matter in heavy-ion collision experiments. These signals are possibly
weakened when the isovector vector interaction is included,
whereas they are slightly strengthened by the isoscalar vector interaction.
However,
the main problem induced by the isoscalar vector interaction is  that the
onset densities of the mixed phase are moved to  higher densities.
Most uncertainty lies in the relevant vector couplings.
%As a further study on this issue, the Fock exchange term will be considered.
%With the Fock exchange term the isospin splitting effect can be
%obtained even with a isospin symmetric Lagrangian.
The planned experiments with
FAIR at GSI-Darmstadt and NICA at JINR-Dubna are expected to provide some hints
on the related study.
%
%{\bf ** The meaning of the following text is not clear to me **}
%In addition, the phase transition density/chemical potential at low
%temperatures in the two-phase model is relatively larger than that
%given by some quark models. This is because the equation of state
%of nuclear matter is much stiffer than that of quark matter,
%which leads to that the pressure can only reach an equilibrium at a
%much higher density for quark matter. Some further physical insights
%will be considered on this aspect in future, such as the
%finite size effect of hadrons.

\begin{acknowledgments}
We would like to thank V.Greco and S.Plumari for stimulating discussions. 

This  work  was supported in part by the National Natural Science Foundation
of China under Grants Nos. 11147144, 11075037, and 10935001.
\end{acknowledgments}

%\end{CJK*}

\begin{thebibliography}{150}
\bibitem{fock}
This is not surprising since in the hadron sector, within
the Relativistic Mean Field approach, the Fock terms also give
a substantial fraction of the effective coupling constants,
\cite{greco01,baran05}.
\end{thebibliography}
\end{document}